\documentclass{webofc}
\usepackage[varg]{txfonts}   

\begin{document}

\title{Hartle-Hawking wave function and large-scale power suppression of CMB\footnote{Proceedings of the 13th International Conference on Gravitation, Astrophysics, and Cosmology \& the 15th Italian-Korean Symposium on Relativistic Astrophysics: A Joint Meeting. Talk on July 7, 2017, Seoul, Republic of Korea}}

\author{\firstname{Dong-han} \lastname{Yeom}\thanks{\email{innocent.yeom@gmail.com}} }

\institute{Leung Center for Cosmology and Particle Astrophysics, National Taiwan University, Taipei 10617, Taiwan}

\abstract{
In this presentation, we first describe the Hartle-Hawking wave function in the Euclidean path integral approach. After we introduce perturbations to the background instanton solution, following the formalism developed by Halliwell-Hawking and Laflamme, one can obtain the scale-invariant power spectrum for small-scales. We further emphasize that the Hartle-Hawking wave function can explain the large-scale power suppression by choosing suitable potential parameters, where this will be a possible window to confirm or falsify models of quantum cosmology. Finally, we further comment on possible future applications, e.g., Euclidean wormholes, which can result in distinct signatures to the power spectrum.
}

\maketitle

\section{Introduction}

In order to resolve and understand the initial singularity of our universe, we need to rely on quantum gravity. There are various approaches, but the most conservative approach is to quantize space and time by following the canonical approach. The corresponding master equation for the wave function of the universe is called by the Wheeler-DeWitt equation \cite{DeWitt:1967yk}.

The Hartle-Hawking wave function \cite{Hartle:1983ai} is a proposal of the boundary condition of the Wheeler-DeWitt equation, where this can be presented by the Euclidean path integral
\begin{eqnarray}
\Psi \left[\partial g_{\mu\nu},\partial \Phi \right] = \int \mathcal{D}g_{\mu\nu}\mathcal{D}\Phi \;e^{-S_{\mathrm{E}} [g_{\mu\nu},\Phi]},
\end{eqnarray}
where $S_{\mathrm{E}}$ is the Euclidean action as a functional of the metric $g_{\mu\nu}$ and the inflaton field $\Phi$. This path integral is over all geometries and fields that have the boundary $\partial g_{\mu\nu}$ and $\partial \Phi$. This path integral can be further approximated by using the steepest-descent approximation, i.e., this integral is presented by the sum over on-shell solutions:
\begin{eqnarray}
\Psi\left[\partial g_{\mu\nu},\partial \Phi\right] \simeq \sum \;e^{-S_{\mathrm{E}}^{\mathrm{on-shell}}},
\end{eqnarray}
where these on-shell solutions are called by instantons. Since several years, dynamical and complexified instantons have been extensively investigated \cite{Hwang:2011mp}.

The Hartle-Hawking wave function not only resolves the initial singularity problem but also gives implications for inflationary cosmology \cite{Hwang:2012bd}. In this presentation, we first revisit the investigations of Halliwell and Hawking \cite{Halliwell:1984eu}. Then by using this, we can investigate physical implications of the Hartle-Hawking wave function.

\section{Halliwell-Hawking revisited}

This section is a brief review of the historical work of Halliwell and Hawking \cite{Halliwell:1984eu}.

We consider the Einstein gravity with a scalar field:
\begin{eqnarray}
S_{\mathrm{E}} = \int \sqrt{g} d^{4}x \left[ \frac{R}{16 \pi} - \frac{1}{2} \nabla_{\mu}\Phi \nabla^{\mu}\Phi - V\left(\Phi\right)\right],
\end{eqnarray}
where $R$ is the Ricci scalar and $V(\Phi)$ is the potential term of the inflaton field $\Phi$. For simplicity, we consider the massive scalar field (with the rescaled scalar field $\phi = \Phi \sqrt{4\pi/3}$)
\begin{eqnarray}
V(\Phi) = \frac{8\pi \sigma^{2}}{3} \left( 1 + m^{2} \phi^{2} \right),
\end{eqnarray}
where $\sigma$ is a constant and $m$ is the mass parameter of the scalar field. It is useful to define the effective Hubble parameter $H_{0} = \sqrt{8\pi/3}$.

In order to consider perturbations to the homogeneous solution, we introduce the following metric and field ansatz ($i, j = 1, 2, 3$ and we assume the summation convention for the spherical coordinate indices $n, l, m$):
\begin{eqnarray}
ds^{2} &=& \sigma^{2} \left[ \left( N^{2} - N_{i}N^{i} \right) d\tau^{2} + 2 N_{i} dx^{i} d\tau + a^{2} \gamma_{ij} dx^{i}dx^{j} \right],\\
N &=& N_{0} \left[ 1 + \frac{1}{\sqrt{6}} g_{nlm}Q_{nlm} \right],\\
N_{i} &=& a \left[ \frac{1}{\sqrt{6}} k_{nlm} \left( P_{i} \right)_{nlm} + \sqrt{2} j_{nlm} \left( S_{i} \right)_{nlm} \right],\\
\gamma_{ij} &=& \left[ 1 + \frac{\sqrt{6}}{3} q_{nlm}Q_{nlm} \right] d\Omega_{3}^{2} + \left[ \sqrt{6} b_{nlm} \left(P_{ij}\right)_{nlm} \right. \nonumber \\ 
&& \left. + \sqrt{2} c^{o}_{nlm} \left( S_{ij}^{o} \right)_{nlm} + \sqrt{2} c^{e}_{nlm} \left( S_{ij}^{e} \right)_{nlm} + 2 d^{o}_{nlm} \left( G_{ij}^{o} \right)_{nlm} + 2 d^{e}_{nlm} \left( G_{ij}^{e} \right)_{nlm} \right],\\
\Phi &=& \sqrt{\frac{3}{4\pi}} \phi + \frac{3\pi}{2} f_{nlm}Q_{nlm},
\end{eqnarray}
where $N_{0}$ is an arbitrary constant, $a = H_{0}^{-1} \sin H_{0}\tau$ and $\phi = \mathrm{const.}$ are the background solutions of the de Sitter space (or an approximate solution of the slowly rolling scalar field), $g$, $k$, $j$, $q$, $b$, $c^{o,e}$, $d^{o,e}$, $f$ are time dependent mode functions and $Q$, $P_{i}$, $S_{i}$, $P_{ij}$, $S_{ij}^{o,e}$, $G_{ij}^{o,e}$ are space dependent expansion basis. The detailed definitions for the expansion basis are in \cite{Halliwell:1984eu}.

In this work, we are focusing on the scalar field perturbations. In the slow-roll limit, the scalar field perturbation $f_{nlm}$ is determined by the following equation \cite{Halliwell:1984eu,Chen:2017aes}:
\begin{eqnarray}\label{eq:fnlm}
\ddot{f}_{nlm} + 3 H_{0} \dot{f}_{nlm} + \left( m^{2} + \frac{n^{2}-1}{a^{2}} \right)f_{nlm} = 0.
\end{eqnarray}
From this, we can calculate the power spectrum $P(n)$ as a function of $n$ (after summing over $l$ and $m$)
\begin{eqnarray}
P(n) = \frac{3n^{2}}{4\pi} \langle f_{n}^{2} \rangle,
\end{eqnarray}
where we simply denote $f_{n}$ since it only depends on $n$.

In order to calculate the expectation value of $f_{n}^{2}$, we need to present the wave function for each perturbation $f_{n}$. By plugging the on-shell condition of $f_{n}$, the wave function will look like
\begin{eqnarray}
\Psi\left[f_{n}\right] \propto \exp\left[ -\frac{1}{2}a^{3} f_{n}\dot{f}_{n} \right].
\end{eqnarray}
There would be many ways to write $\dot{f}_{n}$ as a function of $f_{n}$. However, if we give the Euclidean vacuum condition following Laflamme \cite{Laflamme:1987mx}, one can obtain the proper form:
\begin{eqnarray}
\Psi\left[f_{n}\right] \propto \exp\left[ -\frac{\dot{\tilde{f}}_{n}}{2 \tilde{f}_{n}}a^{3} f_{n}^{2} \right],
\end{eqnarray}
where $\tilde{f}_{n}$ and $\dot{\tilde{f}}_{n}$ are $f_{n}$ and $\dot{f}_{n}$ which are evaluated at the horizon crossing time, respectively. By using this wave form, one can finally obtain a closed form of the power spectrum $P(n)$:
\begin{eqnarray}
P(n) = \frac{3n^{2}}{4\pi} \frac{\tilde{f}_{n}}{2a^{3} \dot{\tilde{f}}_{n}}.
\end{eqnarray}

\begin{figure}[h]
\centering
\includegraphics[scale=0.95]{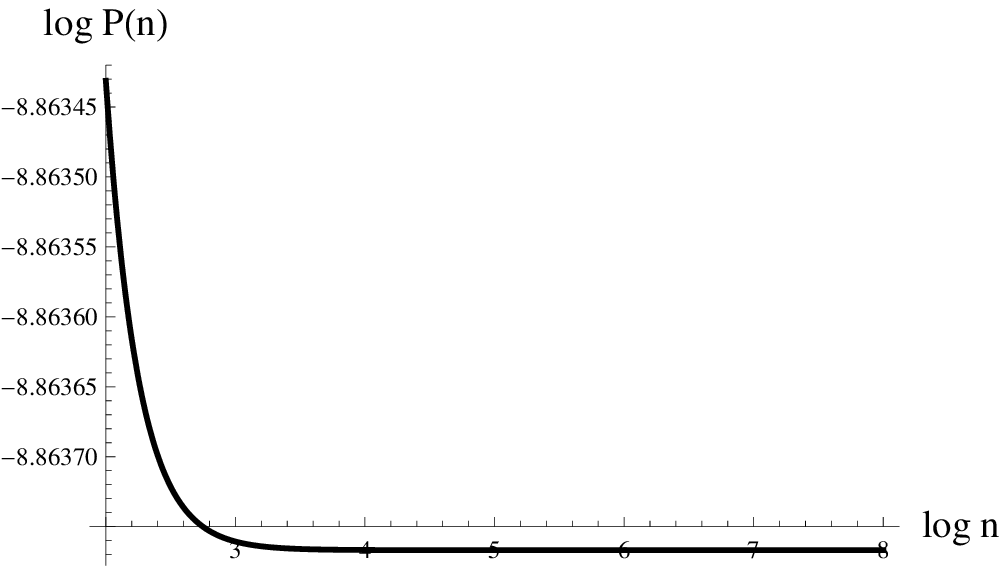}
\includegraphics[scale=0.95]{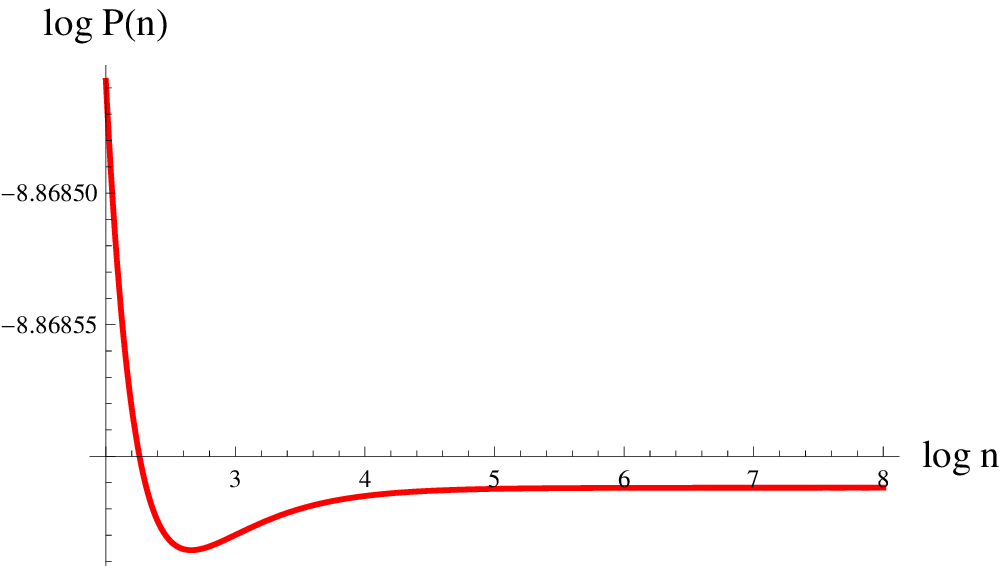}
\caption{The power spectrum $P(n)$ for $H_{0} = 0.01$ and $m = 0$ (upper) and $m = 0.001$ (lower). If the mass term is zero, then the power spectrum is enhanced for small $n$. However, as $m$ increases, there is a possibility that the power spectrum is suppressed for small $n$.}
\label{fig-1}
\end{figure}

\section{Suppression of CMB power spectrum}

By solving equation (\ref{eq:fnlm}), we can obtain $\tilde{f}_{nlm}$. There are two linearly independent solutions, but we need to impose the correct boundary condition at $\tau = 0$. The physically viable solution can be presented by the following analytic form \cite{Laflamme:1987mx}:
\begin{eqnarray}
\tilde{f}_{n} = D \left(z-z^{2}\right)^{(n-1)/2} {}_{2}F_{1} \left(n-\nu,n+\nu+1;n+1;z\right),
\end{eqnarray}
where
\begin{eqnarray}
z &=& \frac{1-\cos H_{0}\tau}{2},\\
\nu &=& -\frac{1}{2} \pm \sqrt{\frac{9}{4} - \frac{m^{2}}{H_{0}^{2}}},\\
D &=& \frac{2^{n+1} \Gamma (1 + n/2 + \nu/2)}{n! H_{0} \Gamma(1 - n/2 + \nu/2)}.
\end{eqnarray}
The time parameter $\tau$ will be Wick-rotated to $\tau = \pi/2H_{0} + it$. The function $\tilde{f}_{n}$ should be evaluated at the horizon crossing time, i.e., $n = aH = \dot{a} = \sinh H_{0}t$ which is the analog relation of $k/aH \simeq 1$ for the flat topology case.

The final result of the power spectrum $P(n)$ is plotted in figure~\ref{fig-1} (more detailed discussions are in \cite{Chen:2017aes}). If the mass term is zero, then the large-scale power spectrum is enhanced (upper), which is consistent with quantum field theoretical expectations \cite{Starobinsky:1996ek}. However, as one turns on the mass term, the large-scale power spectrum can be partly suppressed (lower). Of course, the detailed choices of parameters depend on the detailed inflation model, but it is enough to show that quantum cosmological calculations based on the Hartle-Hawking wave function can explain the large-scale power suppression. Note that for both cases, the power spectrum is scale-invariant for small-scales as we expected.

\section{Future perspectives}

Up to now, the observed CMB data has a tension with the theoretical expectations (based single field inflation and the $\Lambda$CDM model), especially for large length scales \cite{Aghanim:2015xee}. It is not clear whether this tension will be confirmed to be real or not for future experiments. However, if it is the case, then this tension with theoretical expectations is a good window to test quantum gravity.

In this paper, we show that the Hartle-Hawking wave function can easily explain this power suppression behavior. The Hartle-Hawking wave function is not the only way to explain the large-scale power suppression (e.g., \cite{Ashtekar:2016wpi}), but we can say that this approach is very well-established and conservative compared to the others.

We can extend this investigation to the other inflationary models \cite{Hwang:2013nja} and also improve approximations for dynamics of the inflaton field \cite{Hwang:2012mf}. Perhaps, we may study not only compact instantons but also non-compact instantons, e.g., Euclidean wormholes \cite{Chen:2016ask}. In the end, we hope that future experiments can confirm or falsify several models of quantum cosmology. Then, quantum gravity will be established on the experimental ground.

\newpage

\section*{Acknowledgment}

The author would like to thank Pisin Chen and Yu-Hsiang Lin for the stimulated collaborations of this project. The work is supported by Leung Center for Cosmology and Particle Astrophysics (LeCosPA) of National Taiwan University (103R4000).

\end{document}